\documentclass[sigconf]{acmart}

\usepackage{booktabs}
\usepackage{enumitem}
 \usepackage{xcolor}
\makeatletter
\let\footnotetextcopyrightpermission\@gobble
\makeatother

\acmDOI{}
\acmISBN{}
\acmConference[]{}{}{}

\AtBeginDocument{%
  }
\usepackage{multirow}
\usepackage{xcolor}
\usepackage{xparse}
\NewDocumentCommand{\new}{+m}{{\color{blue}#1}}

\begin{document}

\settopmatter{printacmref=false}

\title{AI Adoption Across Mission-Driven Organizations}

\author{Dalia Ali}
\email{dalia.ali@tum.de}
\affiliation{%
  \institution{Technical University of Munich}
  \city{Munich}
  \country{Germany}
}

\author{Muneeb Ahmed}
\email{muneeb.ahmed@tum.de}
\affiliation{%
  \institution{Technical University of Munich}
  \city{Munich}
  \country{Germany}
}

\author{Hailan Wang}
\email{hailan.wang@tum.de}
\affiliation{%
  \institution{Technical University of Munich}
  \city{Munich}
  \country{Germany}
}

\author{Arfa Khan}
\authornote{Both authors contributed equally to this research.}
\email{arfa.khan@tum.de}
\affiliation{%
  \institution{Technical University of Munich}
  \city{Munich}
  \country{Germany}
}

\author{Naira Paola Arnez Jordan}
\authornotemark[1]
\email{naira.arnez@tum.de}
\affiliation{%
  \institution{Technical University of Munich}
  \city{Munich}
  \country{Germany}
}

\author{Sunnie S. Y. Kim}
\authornote{Currently at Apple.}
\email{sunniesuhyoung@princeton.edu}
\affiliation{%
  \institution{Princeton University}
  \city{Princeton}
  \state{New Jersey}
  \country{USA}
}

\author{Meet Dilip Muchhala}
\email{mmuchhala@wwfint.org}
\affiliation{%
  \institution{WWF}
  \city{Mumbai}
  \country{India}
}

\author{Anne Kathrin Merkle}
\email{amerkle@wwfint.org}
\affiliation{%
  \institution{WWF}
  \city{Berlin}
  \country{Germany}
}

\author{Orestis Papakyriakopoulos}
\email{orestis.p@tum.de}
\affiliation{%
  \institution{Technical University of Munich}
  \city{Munich}
  \country{Germany}
}

\renewcommand{\shortauthors}{Ali et al.}
\begin{abstract} 
Despite AI’s promise, little is known about how mission-driven organizations (MDOs) adopt AI. We conducted semi-structured interviews with 15 experts and leaders within environmental, humanitarian, and development organizations across the Global North and South. We find that MDOs adopt AI selectively, with uses in content creation and data analysis, while maintaining human oversight for mission-critical applications. In case of conflicts between the organization's values and efficiency, participants consider sustainability, neutrality, and community trust as constraints that could limit or prevent AI adoption. Individual adoption of AI is not translated into organizational practice due to fragmented data infrastructure, lack of in-house expertise, inertia, ethical challenges, and dependency on vendors. Participants expected AI use in MDOs to involve institutional control over infrastructure, support for missions, and human-centered human-AI collaboration. We contribute an analysis on how MDOs manage AI adoption, identify barriers to responsible deployment, and provide recommendations for designing mission-aligned AI systems.

\end{abstract}


\maketitle

\begingroup
\small
\noindent\textbf{ACM Reference Format:}\\
Dalia Ali, Muneeb Ahmed, Hailan Wang, Arfa Khan, Naira Paola Arnez
Jordan, Sunnie S. Y. Kim, Meet Dilip Muchhala, Anne Kathrin Merkle,
and Orestis Papakyriakopoulos. 2026. \textit{AI Adoption Across
Mission-Driven Organizations}. Preprint submitted and currently under review.
\par
\endgroup

\section{Introduction} 
\label{sec:intro}

Mission-driven organizations (MDOs)\footnote{We use the abbreviation MDOs to refer to mission-oriented organizations whose primary aim is social impact and not profit. This category of organizations includes international NGOs, UN bodies, humanitarian organizations, conservation organizations, and development agencies.} are at the core of the solution for global problems such as disaster management, biodiversity preservation, refugee management, and poverty alleviation. These organizations are increasingly being recognized as key players in the domain of AI for Social Good (AI4SG).\footnote{\emph{AI for Social Good (AI4SG)} is defined as ``an effort to apply AI technologies in ways that support the achievement of the UN Sustainable Development Goals in a socially responsible manner.'' ~\cite{tomavsev2020ai}.} AI is expected to support such organizations in analyzing data, disaster prediction, and decision making in critical situations~\cite{vinuesa2020role, unesco2021ai, un2023ai}.

Prior work suggests that AI adoption in MDOs is shaped by different organizational priorities than in commercial organizations. While commercial organizations often focus on cost, performance, or gaining competitive advantage, MDOs must also consider commitments like neutrality, accountability, environmental responsibility, and building trust with the communities they serve~\cite{davenport2018, schiff2021ai, anzivino2026surfing, coppi2021explicability}; these commitments can constrain technology decisions.

Even though AI for Social Good is getting more attention, we still do not know much about how MDOs actually decide to adopt AI. Most research focuses on specific examples like crisis prediction or environmental monitoring, or on the technology itself, rather than on how these organizations assess, limit, or sometimes choose not to use these tools ~\cite{pantiris2025enhancing,efthymiou2023role,goldkind2025ai,faruq2024ai}.

In the same way, most studies on technology adoption in areas like Human-Computer Interaction (HCI) and Computer-Supported Cooperative Work (CSCW) have focused on commercial and government organizations; therefore, MDOs have not received as much attention~\cite{przegalinska2024collaborative,kim2024public,zhang2025data}. This is important for HCI because MDOs have public-interest commitments that affect how work and responsibility are shared between people and AI. As a result, AI systems need to be evaluated not just for how well they work, but also for how well they fit with ethical, institutional, and mission-related values. There is still little real-world evidence on how these commitments affect the way MDOs adopt, limit, or use AI.

To address this gap, we examine how MDOs approach AI adoption in practice, and we specifically explore:

\vspace{3pt}

\begin{itemize}[noitemsep, topsep=0pt]

\item \textbf{RQ1}: In which application domains are MDOs currently incorporating AI across their operations?

\item \textbf{RQ2}: What are the mission and context-specific barriers that MDOs face in adopting AI?

\item \textbf{RQ3}: How do MDOs envisage the future of mission-aligned uses of AI?

\end{itemize}

\vspace{6pt}

To answer these questions, we conducted semi-structured interviews with 15 AI experts and technology leaders in MDOs across organizations and regions in Global North and South. Our study examines AI adoption through the perspectives of those who design, implement, and govern these systems in real-world organizational contexts. \textbf{Through our research, we found:}

\begin{itemize}

\item \textbf{Current AI practices}: AI adoption is more common in internal operations and data analysis, while the implementation of mission-critical AI is limited and takes place only as part of pilot projects with extensive human oversight.

\item \textbf{Barriers to adoption}: MDOs face five interconnected adoption barriers, including implementation gaps, institutional inertia, ethical challenges, data governance challenges, and vendor dependency, which prevent the scaling of pilots into actual practice.

\item \textbf{Future outlook}: MDO  technology leaders imagine AI adoption within MDO contexts as being centered around sovereignty, mission amplification, and human-centric models of AI adoption.

\end{itemize}

\textbf{Our contributions to the HCI literature are as follows:} (1) We provide an empirical account of how MDOs allocate work and responsibility between humans and AI under real-world ethical and operational constraints, using AI mainly to support routine and analytical work while retaining human oversight in consequential tasks; (2) We identify structural, institutional, and mission-specific conditions that shape when AI is adopted, constrained, or rejected, showing how organizational context influences human-AI collaboration in high-stakes social impact settings; and (3) We identify design recommendations for human--AI systems in MDOs that allow organizations to define where AI can be used and where human judgment must remain, in line with their mission commitments.

\section{Related Work}

\subsection{Technology Adoption in Mission-Driven Organizations}  Over the last two decades, MDOs have extended the use of digital technologies. The usage in the early 2000s was based on pragmatic and context-bound application rather than strategic transformation of their operations via digital technologies \cite{voida2011homebrew, lister2003ngo}. It has been empirically found that MDOs used digital technologies for the implementation of operational tasks, including volunteer management, communication within organizational networks \cite{voida2011homebrew}, fundraising activities, and service delivery management \cite{voida2011shapeshifters}, along with service quality and operation improvements \cite{evans2010training}. 

Regardless of these uses, most organizations lack the proper IT infrastructure to support their information work \cite{voida2011homebrew}. Also, low budget, lack of availability of in-house IT personnel, and absence of training, led many organizations to depend on an improvised, ad hoc assemblages of spreadsheets, email software, paper-based documentation, and occasionally a small-scale database application, resulting in fragmented, repetitive, and isolated data repositories \cite{voida2011homebrew, voida2011shapeshifters}.

Technology adoption was typically externally driven \cite{lister2003ngo} and reactive, requiring MDOs to adapt their work and technology practices to shifts in the broader ecosystem of public, private, and community sectors \cite{voida2011shapeshifters}. Following this trend, several studies found technology adoption to be uneven and dependent on short-term funding cycles rather than integrated into long-term organizational strategies \cite{burt2000information, voida2011homebrew, bopp2017disempowered}.   Recent years have seen an upsurge in the adoption of technology by MDOs in terms of both scope and sophistication. Prompted by the need to be efficient, personalized, make better decisions, and have more impact \cite{claisse2025exploring}, the use of cloud computing \cite{dinata2021cloud}, social media platforms \cite{mcnutt2018technology}, chatbots \cite{dube2024factors, cheng2025leveraging}, and more recently, AI tools \cite{lee2017human, alsolbi2022analyzing, claisse2025exploring, ciolfi2025doing, mcnutt2018technology}. 

Hence, these advances have enabled MDOs to automate various tasks \cite{bhatnagar2025bridging}, including the analysis of donor behavior, improvement of strategic resource management \cite{alsolbi2022analyzing}, personalization of donor engagement \cite{efthymiou2023artificial}, optimization of operational efficiency, and enhancement of stakeholder engagement \cite{faruq2024ai}.  AI adoption is still inconsistent and heavily dependent on external funds and partnerships \cite{bopp2017disempowered, godefroid2024identifying}. While large, well-funded nonprofit organizations can adopt more sophisticated technologies and use them strategically, smaller ones are forced to settle for marginal changes \cite{mcnutt2018technology, bopp2017disempowered}. Prior works show that such behavior is dictated by existing structural constraints. 

Limited IT budget and funders' tendency to prefer supporting projects with ``low overhead" make it difficult to obtain the funds for basic IT infrastructure or even its maintenance \cite{godefroid2024identifying}. Consequently, many nonprofit organizations cannot afford to hire in-house IT expertise and have to work with external third-party platforms, which might be sufficient for management purposes but not aligned with program design or case management practices, making employees adopt these tools or come up with their own solutions \cite{godefroid2024identifying, claisse2025exploring}. Whereas such limitations determine the use of digital technologies generally, there is a unique set of considerations posed by AI, including its increased dependency on data, ethical concerns, and the need for more stringent governance \cite{hanna2025ethical, ntoutsi2020bias}. 

This explains why traditional models of technology adoption in MDOs do not fully account for the current dynamics of AI, making the concept of adoption a good approach to understand how organizations embrace or reject AI systems. On the basis of such dynamics, modern research underscores that AI comes with more than just benefits; it creates tensions around ethics, equity, and governance. \looseness=-1

\subsection{AI Adoption and Tensions in Mission-Driven Organizations}

AI technology brings considerable operational benefits for MDOs; efficiency gains include reduced response times through predictive analytics and improved resource allocation \cite{BajajAIbenefits, HLCMTFAI2024OperationalAI}. For example, UNHCR implemented Project Jetson to predict displaced populations in Somalia using diverse data sources \cite{HLCMTFAI2024OperationalAI, HoffmannProjectJetson}. Similarly, the UN OCHA piloted anticipatory action in Bangladesh, using forecasting-based finance to secure \$5.2 million in CERF funds two weeks before the 2020 monsoon flood peak. This enabled OCHA to assist over 220,000 people before the flooding and more than 320,000 in total, at about half the cost per person compared to the 2019 rapid-response intervention \cite{UNOCHA2020BangladeshAAPilot}. 

However, adopting AI in MDOs presents challenges beyond general technology implementation. Besides common issues like cost, infrastructure, and training, AI systems raise unique concerns about privacy, security, ethics, fairness, and trust because of their opaque, adaptive, and data-intensive nature \cite{raj2025artificial, schiff2021ai, SophiaEthicalAI, rotter2025ai,dube2024factors}.

These challenges become clear when biased AI systems are used in humanitarian or aid settings, as they can reinforce inequality while aiming for efficiency \cite{goldkind2025ai}. Broadly, bias has been documented in commercial facial analysis systems; for instance, gender classification systems showed substantially higher error rates for darker-skinned women \cite{BuolamwiniGebru2018GenderShades}, and people with disabilities \cite{WhittakerEtAl2019DisabilityBiasAI}. This increases discrimination against groups that are already vulnerable \cite{pizzi2021ai}. The tension between efficiency and fairness is especially strong in algorithmic aid allocation, where MDOs must balance quick responses with fair treatment for all demographic groups \cite{beduschi2022harnessing}.

Data analytics capabilities also present similar dualities. AI-driven crisis mapping and social media analytics can improve intervention quality, as shown by the World Food Program's HungerMap, which monitors food security \cite{Herteux2024ForecastingTrends}. However, serious privacy concerns arise, especially regarding 'surveillance humanitarianism,' where biometric data is collected without consent as a condition for accessing services \cite{KerenSurveillanceHumanitarian}. Therefore, MDOs must balance program improvement with protecting beneficiaries' privacy rights.

Environmental considerations further complicate AI adoption. While AI supports sustainable development through efficient resource allocation and climate modeling, its substantial energy use creates significant trade-offs \cite{AIandEnvironment, TradeoffsAIandEnv}. Global electricity consumption by data centers is expected to more than double by 2030 \cite{IEA2025EnergyAI}, and large amounts of water are needed for cooling, which is especially problematic in water-scarce regions where humanitarian operations occur \cite{Barringer2025ThirstyPowerWater}.

Considering these challenges, very little research has been carried out on AI adoption from a mission-driven perspective. The existence of this gap, together with the conflict between efficiency and equity, requires MDOs to make use of AI's potential in a strategic way while still keeping the human-centered approach at the heart of their missions. Therefore, AI adoption has to be seen as a socio-technical process in which organizations negotiate accountability, legitimacy, and core values.

\subsection{Mission-Driven Organizations from a Socio-Technical Perspective} 

Studying MDOs in terms of a socio-technical perspective is crucial, since the adoption of technology cannot be understood in a purely technical process. As illustrated in the previous subsection, the application of AI technologies promises to yield transformative impacts in the domain of humanitarianism, but at the same time brings about tensions around issues of equity, privacy, and environmental sustainability. 

Such tensions get amplified across sectors by the representation of automation through overly idealized imaginaries that obscure the politics and structural dependencies of organization  \cite{gomez2025does, baur2023inserting}, while AI adoption surfaces persistent value conflicts between efficiency and accountability, autonomy and justice \cite{madan2023ai}. Tensions also emerge in MDOs within larger organizational ecologies due to competing stakeholder pressures, accountability systems, and resource limitations affecting technology integration.

MDOs operate within complex socio-technical systems that are shaped by overlapping and often conflicting expectations from donors, governments, and the communities they serve \cite{bopp2020voices}. They exist in dense inter-dependencies with public, private, and community sectors (e.g., policy requirements and funding dependencies \cite{voida2011shapeshifters}; the adoption of business-oriented data systems that evaluate efficiency and performance using metrics that often do not correspond to mission-driven work \cite{bopp2017disempowered}; and reciprocal reliance between nonprofits and local volunteers or clients that shapes everyday technology use \cite{voida2011homebrew}), influencing not only the nature of their work but also the pathways of technology integration \cite{voida2011shapeshifters}. 

Balancing ``upward'' accountability to donors and governmental agencies while maintaining ``downward'' accountability to the communities they serve, technology often becomes a mediator in these negotiations.  As a result, the pressures of upward accountability often push MDOs toward adopting digital systems that primarily serve reporting, auditing, and donor-facing transparency, rather than value-aligned systems that improve mission-driven work \cite{lister2003ngo, voida2011shapeshifters, bopp2017disempowered}.

Research on human-computer interaction reveals that such systems are generally developed under significant resource constraints, resulting in ``homebrew'' technology assemblages  \cite{voida2011homebrew}, which means that they are essentially improvised systems made up of a combination of tools used for essential tasks like coordinating volunteers. Such research emphasizes the need for information system designs that account for stakeholders’ values, maintain meaning and fairness, and promote social interactions apart from being efficient  \cite{lee2017human, bopp2017disempowered, bopp2020voices}. Scholars have advised against techno-solutionism that forces nonprofit organizations to embrace technology quickly without matching tools with organizational logic and values, thus disempowering the organizations \cite{harmon2017design, mcnutt2018technology}.

In response, participatory and co-design approaches have been proposed to align technology with organizational missions and capacities. For instance, \citet{ciolfi2025doing} present AymurAI, a feminist AI tool co-designed with Latin American justice actors to address gender-based violence data gaps, while \citet{lee2017human} demonstrate how algorithmic allocation in food rescue can be designed to reflect fairness and community values. \citet{bhatnagar2025bridging} further argue that HCI-informed frameworks are needed to support responsible AI adoption in humanitarian practice, attending to risks, organizational readiness, and collaboration. \looseness=-1

Previous research has found that decisions related to AI4SG collaborations can be driven by the priorities of both funders and AI developers \cite{lin2024come}, while collaborations between nonprofits and technology firms are characterized by differences in organizational norms and  power imbalances \cite{erete2025towards}. Our study builds on these by investigating how MDO practitioners describe the role of organizational capacity and mission commitments in decisions about whether and under what conditions to use AI. More specifically, our focus is on distinguishing between limited deployment due to organizational capacity constraints, deliberate rejection of particular applications, and adoption conditional on requirements such as data control and human oversight.

\section{Methods}

To investigate the use of AI in MDOs, we employed the Goal-Question-Metric (GQM) framework~\cite{basili1994goal,wohlin2012experimentation}. The framework guided the development of three research goals and their corresponding research questions (Table~\ref{tab:study-framework}). We then conducted semi-structured interviews with 15 practitioners from humanitarian, environmental, and development organizations to address these questions. All research procedures were approved by the Institutional Review Board (IRB) of the corresponding authors' institution.

\begin{table*}[htbp]

\centering

\renewcommand{\arraystretch}{0.5}

\caption{Our study framework: Goals, Questions, and Analytical Focus}

\Description{\new{

The above table illustrates our study framework, which comprises four themes. These themes are Current AI Practices, Challenges in Adoption, Future Outlook, and Recommendations. As illustrated in the table, the rows represent our Goals, Questions, and Analytical Focus. Our Goals illustrate the objective of each theme, while the Questions represent the study questions for each theme. Analytical Focus, on the other hand, illustrates the analysis techniques applied. As demonstrated in the table, our study framework comprises four columns and three rows.

}}

\centering

\label{tab:study-framework}
\renewcommand{\arraystretch}{1.15}
\small
\begin{tabular}{@{}p{1.4cm}p{3.9cm}p{4.2cm}p{4.3cm}@{}}

\toprule
\textbf{} & \textbf{Current AI Practices} & \textbf{Barriers in Adoption} & \textbf{Future Outlook} \\

\midrule

\textbf{Goals}

& Analyzing how MDOs  deploy AI across their operations

& Identifying the organizational, ethical, and infrastructural barriers that constrain adoption

& Exploring practitioners' visions of AI's role in strengthening missions while preserving values \\

\midrule

\textbf{Questions}

& In what application areas are MDOs currently adopting AI across their operations?

& What mission- and context-specific barriers do MDOs face in adopting AI?

& How do MDOs envision mission-aligned and values-consistent future uses of AI? \\

\midrule

\textbf{Analytical Focus}

& Inductive thematic coding of interview data to identify domains of AI use and practices

& Inductive clustering of barrier-related data into higher-order categories through constant comparison

& Narrative and thematic coding of future reflections, synthesized into organizational outlooks \\

\bottomrule

\end{tabular}

\end{table*}

\subsection{Participant Recruitment and Selection}

We targeted participants who are experts in AI and senior technology professionals involved in AI adoption within MDOs . We considered differences in organizational hierarchy, domain focus, and geographical context. Participants were selected based on their technical or strategic roles in AI adoption or governance, substantial professional experience, and direct involvement in their organizations' AI initiatives.

Our participant selection strategy involved a wide professional network with various professionals from diverse local and international NGOs, technology consortium, and intergovernmental organizations who are actively engaged in AI for social good initiatives. Our participants were from diverse geographical settings across Europe and North America (Global North) and South Asia (India), East Asia (Hong Kong), and the Middle East and North Africa (MENA) region, representing diverse geographical settings from Global North and Global South (Table ~\ref{tab:participants}).

\begin{itemize}

\item \textbf{Strategic level (n=6):} Global heads of AI/data, digital transformation directors, and senior advisors for AI governance and strategy.

\item \textbf{Program level (n=5):} Regional transformation leads, senior project managers, and technical leads for AI adoption in their organizations.

\item \textbf{Implementation level (n=4):} Applied data scientists, innovation managers, and specialists for AI adoption in their organizations.

\end{itemize}

The final sample included participants from environmental (n=11), humanitarian (n=3), and development (n=1) organizations. In light of this imbalance, we treat cross-sectorial comparisons as exploratory but not representative of differences between MDO sectors.

\subsection{Study Procedure}

\begin{figure*}[htbp]
    \centering
    \makebox[\textwidth]{\includegraphics[width=15cm, height=13cm]{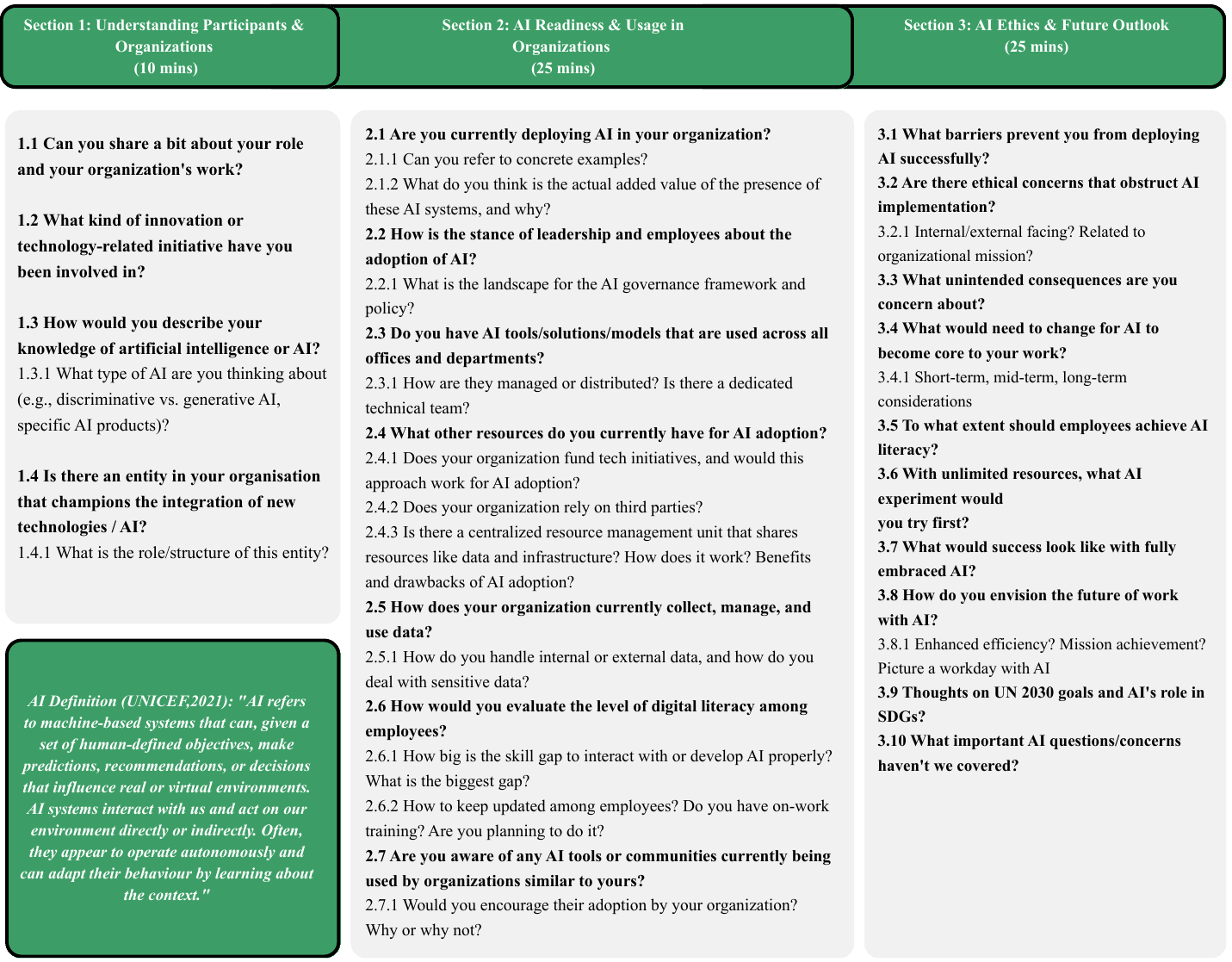}}
    \Description{\new{ The figure presents the full interview guide structured into three sections. Section 1, titled ``Understanding Participants and Organizations'' (10 minutes), includes introductory questions about the participant’s role, their organization’s mission, prior innovation or technology-related initiatives, and their general understanding of artificial intelligence. This section also contains the UNICEF (2021) definition of AI, which describes AI as machine-based systems that make predictions, recommendations, or decisions based on human-defined objectives and that may appear autonomous by adapting their behavior to context. The section concludes with questions about whether the organization has an entity that champions new technologies and how that entity is structured. 
    
    Section 2, titled ``AI Readiness and Usage in Organizations'' (25 minutes), covers questions about whether AI is currently deployed, concrete examples of use, perceived value of AI systems, leadership and employee attitudes, existing governance frameworks, organization-wide tools, availability of technical teams, funding structures, reliance on third parties, data management practices including handling of sensitive data, digital literacy and skill gaps, training practices, and awareness of AI tools used in similar organizations. 
    
    Section 3, titled ``AI Ethics and Future Outlook'' (25 minutes), addresses barriers to adoption, ethical concerns, potential unintended consequences, the conditions needed for AI to become core to organizational work, appropriate levels of AI literacy, desired experiments if resources were unlimited, the vision of a future workday with AI, the role of AI in achieving the UN 2030 Sustainable Development Goals, and an open prompt for additional concerns or overlooked issues. }}

    \caption{Semi-structured interview instrument comprising three sections: (1) Understanding participants and organizations, (2) AI readiness and organizational use, and (3) Ethical and future considerations. The instrument includes 30 questions and sub-questions designed to explore organizational AI adoption patterns, barriers, and future perspectives.}
    \label{fig:interview_instrument}
\end{figure*}

Each interview began with obtaining informed consent, which included an explanation of the study’s aims, data handling procedures, and confidentiality measures. Participants were notified that interviews would be recorded and analyzed. The interviews had three sections; see Figure~\ref {fig:interview_instrument} for the full interview instrument. To establish a shared understanding, we provided participants with the following definition of AI proposed by UNICEF (2021)~\cite{unicef2021policy}: \textit{``AI refers to machine-based systems that can, given a set of human-defined objectives, make predictions, recommendations, or decisions that influence real or virtual environments.''}
 
\textbf{1. Understanding Participants and Organizations.} This section of the interview sought to understand the participants' roles, organizational settings, and their previous initiatives in innovation. Examples of interview questions include: \textit{``Can you share a bit about your role and your organization’s work?''} and \textit{``Is there an entity in your organization that champions the integration of new technologies or AI?'} \looseness-1

\textbf{2. AI Readiness and Organizational Use.} In this section, specific issues related to the adoption, governance, and use of AI in organizations were explored. Some questions used in the prompts included: \textit{``Are you currently deploying AI in your organization? Can you refer to concrete examples?''}, \textit{``How is the stance from leadership and employees regarding adoption?''}, and \textit{``What resources, such as technical, financial, or partnerships, support AI adoption in your organization?''}. Other questions pertained to issues like data governance, digital literacy, and centralized or decentralized resource management.

\textbf{3. Ethical and Future Considerations.} In the final section, questions focused on the barriers to AI adoption, ethics, and future visions. Some questions posed to the respondents included: \textit{``Are there ethical concerns that obstruct AI implementation in your institution?''}, \textit{``What unintended consequences do you worry about?''}, and \textit{``If your organization had unlimited resources, what would you try first?''}. To encourage respondents to think about future visions of AI use in their organizations, they were asked to ``imagine your future workday with AI`` and ``link AI’s potential to the UN SDGs''

\begin{table*}[htbp!]

\centering

\caption{Participant demographics and organizational context}

\Description{ This table summarizes the demographic and organizational characteristics of the 15 participants interviewed. It lists each participant’s ID, their role or position, their primary work area, the type of organization they work for, and the global region in which they are based. Participants come from international NGOs, local NGOs, IGOs, and technology consortia, with work spanning environmental, development, and humanitarian domains. Regions represented include Europe, North America, South Asia, the Middle East and North Africa, and Asia-Pacific. The table contains 15 rows of participants and five columns describing their profiles. }

\label{tab:participants}
\renewcommand{\arraystretch}{1.25}
\small

\begin{tabular}{@{}p{0.5cm}p{4.8cm}p{2.2cm}p{2.7cm}p{3.4cm}@{}}
\toprule

\textbf{ID} & \textbf{Position} & \textbf{Work Area} & \textbf{Organization Type} & \textbf{Region} \\

\midrule

P1 & Chief Regional Advisor & Development & IGO & Middle East \& North Africa (MENA) \\

\midrule

P2 & Global Head Data \& AI & Environmental & International NGO & Europe (Switzerland) \\

P3 & Head Digital Transformation \& Data & Environmental & International NGO & Europe (Switzerland) \\

P4 & Experiments Team Co-Lead & Environmental & International NGO & Europe (UK) \\

P5 & Director of ICT & Environmental & International NGO & Europe (UK) \\

P6 & Data \& Technology Global Lead Scientist & Environmental & International NGO & North America (USA) \\

P7 & Data Scientist & Environmental & International NGO & North America (USA) \\

P8 & Senior Project Manager Counter & Environmental & International NGO & Asia-Pacific (Hong Kong) \\

P9 & Global Program Manager & Environmental & International NGO & Europe (Sweden) \\

P10 & Director & Environmental & International NGO & North America (USA) \\

P11 & Strategic Innovation \& AI Lead & Environmental & International NGO & South Asia (India) \\

P12 & Co-Founder/Director & Environmental & Local NGO & Europe (Greece) \\

\midrule

P13 & Global AI Advisor & Humanitarian & International NGO & Europe (Switzerland) \\

P14 & Director Center for Digital Nonprofit & Humanitarian & Technology Consortium & Europe (Germany) \\

P15 & Director Innovation, Impact \& Information & Humanitarian & International NGO & Europe (France) \\

\bottomrule

\end{tabular}

\end{table*}
\subsection{Conducting and Analyzing Interviews}

We conducted interviews between March and May 2025, each lasting approximately one hour. All interviews were held online via video conferencing and involved two interviewers. We analyzed the data using Braun and Clarke's thematic analysis framework~\cite{braun2006thematic}. To enhance interpretive validity, we triangulated findings by involving multiple authors in coding, validating the codebook with domain experts, and incorporating feedback from interviewees. We organized the interview guide around the research questions, and we developed the subthemes inductively from the data and synthesized recommendations from participants' suggestions.

Two authors independently coded all 15 transcripts; discrepancies were resolved through discussion across three rounds of coding, with the codebook revised after each round based on these discussions; for instance, early codes for 'financial constraints' and 'business case challenges' were merged into a single representative code under Institutional Inertia after discussion clarified they reflected the same underlying barrier (Table~\ref{tab:T1}). A third author reviewed the final codebook and thematic groupings for consistency. 

The analysis yielded four themes: (1) \textit{Current AI Practices}, describing organizational use of AI across operational domains; (2) \textit{Barriers to Adoption}, examining challenges and sources of resistance to AI adoption and implementation; (3) \textit{Future Outlook}, capturing participants’ visions for AI in mission-driven contexts; and (4) \textit{Recommendations}, synthesizing actionable strategies for aligning AI adoption with organizational values and constraints, based on participants’ insights and our analysis. Additional details on the study procedure and analysis methodology are provided in Appendix~\ref{appendix:method}.

\section{Results}\label{sec:results}
\begin{figure*}[htbp]
    \centering
    \makebox[\textwidth]{\includegraphics[width=11cm, height=16cm]{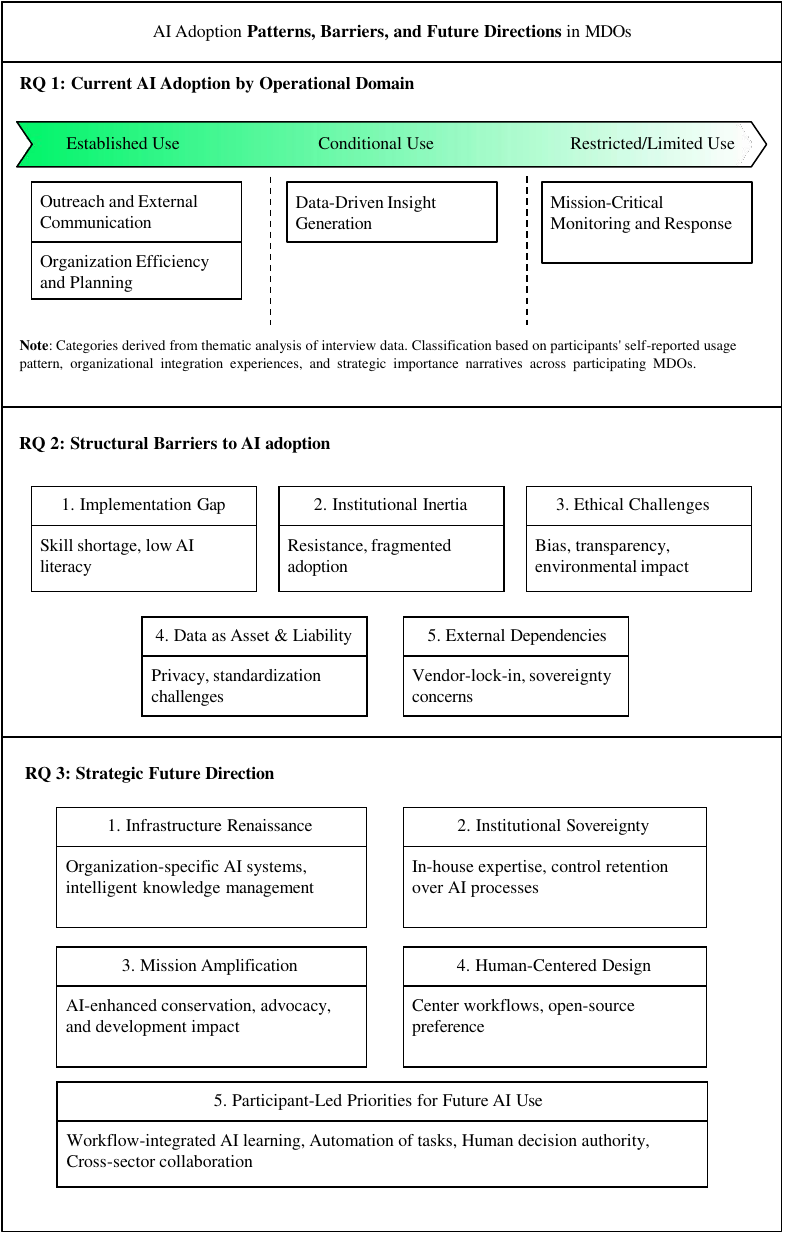}}
    \Description{A diagram divided into three sections summarizing AI-related themes in mission-driven organizations. The first section shows reported patterns of AI use across three categories: established use, conditional use, and restricted or limited use, each with examples of organizational functions. The second section lists five organizational constraints affecting AI adoption, including implementation gaps, institutional inertia, ethical concerns, data challenges, and vendor dependence. The third section presents four future priorities: infrastructure renaissance, institutional sovereignty, mission amplification, and human-centered design.}

    \caption{AI adoption patterns across environmental, humanitarian, and development MDOs, showing reported patterns of use, organizational constraints, and future priorities for human--AI collaboration.}
    \label{fig:resultsec}
\end{figure*}

AI adoption within MDOs is primarily bottom-up, problem-driven, and conditional. MDOs often use AI to reduce administrative burdens and enhance analytical capacity. However, adoption was constrained or rejected where AI challenged organizational missions or the sovereignty of the organization's data; this pattern contrasts with general technology adoption models, which prioritize efficiency and implementation readiness. Additionally, participants believed that AI is a distinct form of technology, as it raises issues related to opaqueness, data dependence, bias, and external control. Figure~\ref{fig:resultsec} provides an overview of these findings.

\subsection{RQ1: Current AI Practices}\label{subsec:rq1}

Participants recognized four main areas in which  AI is currently used in MDOs, which are outreach and external communication, organizational efficiency and planning, data-driven insight generation, and mission-critical monitoring and response. The use of AI was more common in tasks that were already digitized, text-intensive, standardized, and considered relatively low risk. On the other hand, the least use has taken place in tasks where there is a lot of uncertainty about the context, contested data, or direct consequences for vulnerable communities. The primary pattern observed was the degree of alignment between AI capabilities and both the operational and ethical requirements of mission-driven work.

\subsubsection{Outreach and External Communication}\label{subsec:rq1_outreach}

Six participants (\textbf{P2, P3, P7, P10, P11, P14}) mentioned that they are using AI for outreach and external communication activities, including social media content, donor updates, stakeholder summaries, and external-facing reports. In all cases, AI generated initial drafts that they later adapted for particular purposes.

\begin{quote}\emph{
``We use AI for content creation such as writing social media posts, reports,  newsletters, and even donor communications. It ensures uniformity of voice across platforms and languages.'' (\textbf{P11})
}\end{quote}

Participants noted that such efficiencies were dependent on the structure of communication workflows; when donor data or messaging formats were standardized, AI supported in reducing the drafting time. \textbf{P7} described donor-modeling workflows where automated drafts facilitated the scaling of personalized outreach without increasing staff workload.

The participants also emphasized the need for human involvement, especially in cases of multilingual or culturally specific situations; much time would be spent making the adjustment in the tone or terminology, and matching with relationship-building norms. As \textbf{P10} mentioned that contextualization is crucial in interactions with external partners, whereas \textbf{P2}  highlighted that considerable post-editing work is required in multilingual  policy contexts to avoid misinterpretation.

Generally, these applications helped in cutting down the amount of work required for drafting and redirected efforts to verifying and making sure of alignment with organizational values, where credibility and trust are important.

\subsubsection{Organizational Efficiency and Planning}\label{subsec:rq1_efficiency}

Nine participants (\textbf{P2, P3, P5, P7, P9, P10, P11, P14, P15}) discussed the use of AI to enhance MDOs' efficiency in terms of streamlined processes, especially in routine activities like summarization, document processing, report generation, and coordination. 

Participants frequently mentioned tools such as Copilot and ChatGPT to assist with drafting and summarization, and linked these uses to workload pressures. \textbf{P5}, for example, used Copilot to summarize meeting minutes and assist with drafting emails or contracts; \textbf{P7}  used AI in donor-reporting workflows and content tagging pipelines, while \textbf{P9} used it to support the preparation of briefing materials and summaries of technical documents.

\begin{quote}\emph{
``we try to motivate people to use these tools because we see efficiency gains and quality improvements'' \textbf{-P3}
}\end{quote}

Participants emphasized that these benefits depended on organizational infrastructure, \textbf{P3} and \textbf{P15} identified integration challenges caused by fragmented databases, outdated software, and inconsistent data standards. Also staff often had to coordinate outputs across systems, verify results, and adapt AI-generated material to organizational formats. \textbf{P14} similarly noted that differences in digital infrastructure across partner organizations limited automation.

Participants also noted that efficiency gains differ across different parts of the organization, for example, in heavily document-based work departments such as administrative, financial, or legal departments, employees gained the most immediate benefits. By contrast, participants working in humanitarian or conservation operations described fewer improvements, as field-level decision-making often relied on incomplete, fast-changing, or locally specific information. \textbf{P11} and \textbf{P10} further pointed to differences in digital literacy across teams, which affected how easily staff could adopt AI tools. 

Even though AI applications, such as document drafting, summarization, and reporting, are adopted in MDOs, their adoption is shaped by AI-specific constraints, including fragmented data infrastructure, decentralized partner systems, and field-based decision environments, which limit system integration and automation. Thus, AI was used as an auxiliary technology necessitating continuous human coordination, in line with past research on AI in humanitarian and nonprofit contexts~\cite{fernandez2018humanitarian,pu2025challenges}.

\subsubsection{Data-Driven Insight Generation}\label{subsec:rq1_insight}

Seven participants (\textbf{P2, P3, P7, P10, P11, P14, P15}) mentioned the application of AI for the automated analysis of sustainability reports, geospatial information, environmental information, or compliance reports. They use AI to speed up the analysis process while retaining human expertise and judgment.
\begin{quote}\emph{``We worked with [a university] to create an AI system that analyzes sustainability reports… we defined indicators, and the system goes through each one to identify red flags.''\textbf{-P3}
}\end{quote}

Participants shared cases where these systems were applied practically, for example \textbf{P3} highlighted a system jointly developed by university partners which analyzes corporate sustainability reports against predefined environmental indicators, then signals inconsistencies or missing disclosures. Also, \textbf{P3} pointed out that they used an AI system that screens hundreds of building applications against  databases of protected areas and identified cases of construction around water bodies and forests.

However, these systems have limits; participants highlighted that models were often built around established scientific taxonomies, regulatory classifications, or organizational indicators. \textbf{P2} noted that tools trained on standardized global conservation categories struggled to incorporate locally specific ecological knowledge or context from field reports. \textbf{P15} said that systems based largely on Western environmental frameworks could overlook community-based land-use practices, particularly in contested or under-documented areas.

Participants also raised concerns about the transparency and defensibility of these systems. For instance, \textbf{P11} expressed concern over how generative systems arrived at their outputs, thus finding it hard to justify these outputs to stakeholders.  \textbf{P15} pointed out that false positives in politically sensitive contexts, such as land disputes or corporate-government negotiations, could create reputational risks when they conflicted with local perspectives.

These systems were most effective in structured data contexts, while heterogeneous or contested data required substantial human verification, reinforcing AI’s role as support rather than decision-making infrastructure.

\subsubsection{Mission-Critical Monitoring and Response}\label{subsec:rq1_missioncritical}

Five participants (\textbf{P1, P7, P8, P13, P15}) discussed the use of AI for mission-critical monitoring and response operations. For conservation, participants mentioned their use of AI systems for the classification of wildlife images, environmental change detection, and notification for field teams. \textbf{P7} described how computer vision systems are developed with structured species datasets and verified by humans. Humanitarian participant explained that the current AI technology was not suitable since their work involves limited data, quick changes, and political sensitivity. 

\begin{quote}\emph{
``A major concern is losing control over results - missing vulnerable populations or having blind spots that compromise our analysis.'' \textbf{-P15}
}\end{quote}

\textbf{P13} and \textbf{P15} described fragmented information and context-dependent interpretation as barriers to automation. They noted that automated outputs can omit information needed to determine an appropriate response, requiring further human assessment. Prior work likewise identifies reliable data, transparent provenance, and human oversight as requirements in high-stakes settings~\cite{WiltonPark2024,OCHA2024}.

Thus, it can be argued that AI is more readily adoptable in domains with well-structured data, such as conservation, than in humanitarian domains, which are characterized by uncertainty and fragmented information.

\paragraph{Synthesis of RQ1.}
Across these domains, AI adoption in MDOs was not determined primarily by technical capability, but by the degree of alignment between AI systems and mission-specific constraints. Adoption was most successful in contexts characterized by structured data, standardized workflows, and low operational risk, while it remained limited in settings involving contextual uncertainty, contested knowledge, or direct consequences for vulnerable communities.

These findings suggest that AI in MDOs functions less as a general-purpose automation technology and more as a context-dependent augmentation layer, whose applicability is bounded by the ethical, operational, and epistemic conditions of mission-driven work.

\subsection{RQ2: Barriers in Adoption}\label{subsec:rq2}  

To understand the adoption of AI in MDOs, it is necessary to move beyond the technical feasibility of AI and focus on the value systems of the organizations, which affect their resource availability and mission priorities. Our results show that ethics, data, inertness, and vendors can limit adoption in different ways, from preventing projects from moving beyond pilot experimentation to restricting or excluding particular uses.

\subsubsection{The Implementation Gap}\label{subsec:rq2_implgap}  

Out of 15 participants, 7 \textbf{(P2, P3, P7, P10, P11, P14, P15)} reported that there exists an implementation gap between the use of AI in daily life and the underlying capabilities that are necessary to implement reliable AI for MDOs. 

 \begin{quote}\emph{  ``People use ChatGPT every day but do not know how to use it efficiently.'' \textbf{-P11}  }\end{quote}  

Participants explained that this gap reflects broader organizational capacity constraints rather than simple resistance to new technologies. Uneven digital skills, fragmented training opportunities, and heavy workloads leave little time for structured learning or experimentation. As a result, teams often rely on ad hoc knowledge-sharing rather than systematic capacity building. The recruitment barriers also show that the organizations face issues in hiring the necessary workforce, as they are unable to pay salaries as compared to the private sector. \textbf{P15} stated that they are struggling to attract qualified AI engineers.  

The lack of in-house expertise also has an impact on  procurement and implementation support, usually leaving non-specialists to evaluate tools and systems. This causes many projects to fail to scale, whether because of insufficient maintenance capacity or reliance on parallel manual processes, leading to duplication of work. Similar constraints have been documented in nonprofit contexts, where financial and technical limitations hinder sustained AI adoption~\cite{rotter2025ai,godefroid2024identifying}.

The implementation gap results from the combined effects of limited expertise, recruitment challenges, and insufficient organizational capacity, which keep many MDOs in an extended pilot phase where experimentation is frequent but sustained deployment is rare.

\subsubsection{Institutional Inertia}\label{subsec:rq2_inertia}

Eight participants \textbf{(P3, P5, P7, P9, P11, P12, P14, P15)} described forms of organizational inertia that slow AI adoption despite growing interest.   They attributed this inertia to a combination of organizational caution, uneven technical capacity, and governance processes that move far more slowly than AI development cycles. Participants attributed this caution in part to past digital initiatives that failed to deliver expected benefits, shaping more skeptical attitudes toward new technologies.   

\begin{quote}\emph{ ``AI is another buzzword.'' \textbf{-P15} }\end{quote}     

Participants described situations in which technically advanced units experiment with AI tools while other teams continue to rely on paper-based or manual workflows. \textbf{P5} noted that these uneven levels of digital maturity complicate organization-wide adoption because innovations developed in one unit cannot easily be implemented across the entire organization.   

Participants also linked inertia to governance and approval processes that are poorly suited to iterative AI development. Pilots with promising results often stalled in review or procurement cycles, eroding momentum and delaying deployment (\textbf{P11}, \textbf{P3}). Similar dynamics have been documented in NGO contexts, where risk-averse governance slows technology adoption~\cite{godefroid2024identifying}.   

Additionally, the institutional memory of past technological failures also plays a part in this. This is because the participants cited past digital initiatives that failed because of poor oversight and overly ambitious implementation goals. For instance, the 2018 audit of the Uganda refugee registration project highlighted the failure of biometric and digital ID systems, which led to inflated population numbers, donor suspensions, and ultimately forced NGOs in the country to stop operations and rethink distribution strategies \cite{oiosuganda2018}.    

Organizational inertia served as a risk management strategy, with uneven technical capacity, slow governance processes, and recollections of past digital failures reinforcing a preference for cautious experimentation rather than rapid scaling.

\subsubsection{Ethical Challenges}\label{subsec:rq2_ethics}

Nine participants \textbf{(P3, P7, P8, P9, P10, P11, P12, P14, P15)} raised ethical issues surrounding the environmental impact, bias, and the suitability of the use of AI in certain situations. The participants mentioned that there are no adequate measures in place for the organizations to be able to measure the environmental impact of AI systems, for instance, \textbf{P3} and \textbf{P11} noted that there is a lack of information regarding energy consumption, data usage, and carbon emission, making it difficult to verify vendor claims, aligned with prior work on the carbon intensity of AI in cloud computing~\cite{dodge2022measuring}. 

\begin{quote}\emph{
``We are trying to improve sustainability, but the tools we use might impact the environment worse than how much they benefit.'' \textbf{-P11}
}\end{quote}

In addition, participants discussed the possibility of bias since the training of machine learning algorithms on datasets from the West results in poor performance in Global South environments and misinterpretation of multilingual and culturally specific information (\textbf{P8}). There were doubts regarding how reliable the system would be in certain situations.

Furthermore, participants questioned whether the use of AI aligns with organizational values such as neutrality, independence, and communal trust, given the ongoing ethical and environmental challenges associated with AI implementation (\textbf{P15}). Similar challenges have been documented in cases such as the \textit{Huduma Namba} biometric identification system in Kenya, where insufficient safeguards resulted in risks of exclusion and privacy violations~\cite{musoni2023digital}.

Ethical decision-making within MDOs was marked by significant uncertainty since the organizations were required to balance possible benefits against ambiguous environmental, social, and operational risks in the absence of reliable mechanisms for assessing accountability. Participants described a range of outcomes in response to ethical concerns; \textbf{P3} avoided using AI for recruitment and personal data management due to concerns about bias and privacy, and also excluded other potentially valuable use cases with unresolved ethical risks. \textbf{P2} rejected ChatGPT on the basis that it was not open source. 

In other instances , ethical concerns resulted in restricted rather than entirely rejected use. \textbf{P5} preferred Copilot because organizational data remained within SharePoint, while \textbf{P2} declined to use AI-supported decisions when model outputs lacked explainability. These examples illustrate both deliberate non-adoption and conditional adoption.

\subsubsection{Data as Liability}\label{subsec:rq2_data}

Ten participants \textbf{(P3, P5, P7, P8, P9, P10, P11, P12, P14, P15)} described data as both a critical asset for AI and an ongoing organizational liability.

\begin{quote}\emph{
``Data is a nightmare. We generate huge volumes, but we actually have very little usable data because everything is stored in different formats and systems.'' \textbf{-P15}
}\end{quote}
The participants described the current state of disintegrated data as being caused by years of reporting needs and project-based data acquisition. \textbf{P7}, \textbf{P11}, and \textbf{P15} described how many information systems were initially built to meet donor compliance requirements and not to be integrated with other systems. This has resulted in reports, monitoring tools, and field-collected data being spread across different but incompatible systems. Similar patterns have been observed in nonprofit data ecosystems~\cite{achamkulangare2017review}.

Security concerns further constrain data use. Participants emphasized preferences for tools that keep data within organizational infrastructure \textbf{P5} and highlighted challenges in cross-border data sharing due to differing privacy regulations \textbf{P5}, \textbf{P11}. These constraints are particularly critical in sensitive contexts, especially in conflict zones, where data misuse could pose reputational or political risks~\cite{imboden2013infosec}.

Additionally, there is a representation constraint in data. \textbf{P3} and \textbf{P7} explained that some datasets, especially those developed in Western cultures, might not understand local languages or might lack local knowledge. This is in line with data justice in the Global South, which focuses on data representation and how data might limit or reduce the relevance of AI in non-Western countries \cite{itforchange2025,kwet2019digital,lynch2023tears}.

Data functioned as a double-edged condition of AI adoption, it is essential for building AI systems, yet fragmentation, privacy constraints, and representational limits restricted its effective use, making data governance a prerequisite for meaningful adoption.

\subsubsection{External Dependencies}\label{subsec:rq2_dependency}

Nine participants \textbf{(P3, P5, P7, P8, P9, P10, P11, P14, and P15)} reported considerable dependence on external AI companies, primarily due to concerns about vendor lock-in, data ownership, and geopolitical restrictions. In their view, procurement decisions revolved around a few vendors, while the real needs of the organization were often overlooked.

\textbf{P7}, \textbf{P11}, and \textbf{P15} have stated that the lack of internal expertise, combined with the procurement requirements set by donors, prevents them from exploring other options, and therefore, they continue to be locked in by vendors. This aligned with previous findings showing how organizational and financial barriers in nonprofit procurement reduce vendor diversification~\cite{mbiu2015procurement,godefroid2024identifying}.

\begin{quote}\emph{
``It often feels like extracting value from the Global South, similar to colonialism.'' \textbf{-P3}
}\end{quote}

Price constraints further reinforce this dependence, as high-end AI systems are often unaffordable, pushing MDOs toward widely available but poorly governed consumer tools \textbf{P5}, \textbf{P11}. This creates asymmetries in power between technology providers and MDOs.

Also, the participants mentioned the issue of data sovereignty, noting challenges in complying with cross-border privacy regulations and risks associated with storing sensitive beneficiary data in external infrastructures (\textbf{P5}, \textbf{P11}). In reality, these constraints often push MDOs toward large vendors with established compliance tools, even when this conflicts with long-term autonomy~\cite{vannini2020humanitarian,schniederjans2016humanitarian,marelli2020hacking}.

This dependency reflected structural constraints rather than individual choice, as limited technical capacity, restrictive procurement environments, and reliance on global platforms reduced organizational autonomy and bargaining power.

\paragraph{Synthesis of RQ2.}
These barriers show that AI adoption in MDOs is constrained by technical feasibility and the organizational conditions under which AI can be governed legitimately. Implementation gaps and institutional inertia limited the capacity to move beyond experimentation, while ethical uncertainty, data constraints, and vendor dependence raised deeper questions of accountability, sovereignty, and mission alignment. These barriers produced different adoption outcomes since some MDOs remained in experimentation because of limited capacity, others excluded applications over ethical or governance concerns, and others adopted AI only under conditions such as data control and human oversight.

\subsection{RQ3: Future Outlook}\label{subsec:rq3}
Participants envisioned priorities for future AI adoption, which include stronger governance, greater institutional control over data and systems, and Human-AI collaboration. They also preferred open-source or on-premises systems for handling sensitive organizational information, such as data of vulnerable communities they serve.    

\subsubsection{Infrastructure Renaissance}\label{subsec:rq3_infra}

Seven participants \textbf{(P5, P7, P9, P10, P11, P14, P15)} saw the possibility of an Infrastructure Renaissance in which AI systems are expected to bring about a renewal in organizational infrastructure, allowing for a new method of knowledge management and access by changing the way in which knowledge and institutional memory are stored and accessed in the future.  

\begin{quote}\emph{
``Right now people often rely on personal knowledge because no one really knows where things live. Ideally we would have an AI layer that can search across everything we have done as an organization.'' \textbf{-P10}
}\end{quote}

Participants proposed systems that could retrieve and synthesize information from reports, evaluations, and field documentation. \textbf{P11} described an organization-specific system as an ``AI brain'' that could consolidate knowledge accumulated over time; \textbf{P7} and \textbf{P9} expected these systems to support faster decision-making and reduce duplicated work across teams.

They also identified real-time translation as a mechanism for sharing knowledge across country offices. \textbf{P14} and \textbf{P15} anticipated that AI-assisted translation would enhance the accessibility of knowledge produced in one location for teams operating in different linguistic contexts.

Current initiatives by MDOs started to integrate AI systems into digital public infrastructure, for instance, UNDP programs have supported the development of infrastructure for data sharing and the establishment of governance mechanisms to enhance transparency \cite{UNDP2023DPI}. Consequently, participants envisioned AI as an integral component of organizational knowledge infrastructure, rather than merely a tool for individual productivity.

\subsubsection{Institutional Sovereignty}\label{subsec:rq3_sovereignty}

Six participants \textbf{(P3, P5, P7, P11, P14, P15)} wished for their organizations to have more control over the systems employed in their main activities. Their worry was not merely about deciding whether to develop AI themselves or buy it from a vendor; it was about whether the organization could determine where sensitive data were processed, how the models were used, and who was allowed to examine their outputs.

\begin{quote}\emph{
``AI will affect core processes. It cannot remain something that sits completely outside the organization.'' \textbf{-P3}
}\end{quote}

From the interview, it was envisioned that in the future, institutions will have developed the capability to control how AI interacts with sensitive program data. \textbf{P5} stated that control over the environment in which data is processed is a crucial first step toward sovereignty, while \textbf{P11} stated that systems governed internally are more transparent and auditable. \textbf{P14} stated that developing sovereignty over cloud infrastructure might actually be an important component for organizational resilience when operating in politically sensitive environments.

Participants envisioned that in the future, institutions will have developed governance mechanisms that clearly establish boundaries for automated decision-making. \textbf{P3}, \textbf{P5}, \textbf{P11}, and \textbf{P15} stated that institutions need mechanisms such as model registries, internal audit trails, and oversight committees to ensure that AI systems remain accountable to institutional values and humanitarian principles.

\subsubsection{Mission Amplification}\label{subsec:rq3_mission}

Through the interviews, nine participants \textbf{(P3, P5, P7, P8, P9, P10, P11, P14, P15)} described AI as having the potential to amplify the results of missions related to conservation, humanitarian relief, and sustainable development.

Participants described AI systems that can amplify the results of environmental monitoring and risk assessment. \textbf{P10} described AI applications that can automatically analyze camera trap images of different species, while \textbf{P2} described the use of satellite imagery and patterns in large datasets to support forest and wildlife protection. These camera-trap applications are similar to the Wildlife Insights platform, which uses computer vision to automatically classify images of biodiversity at scale \cite{wildlifeinsights2025}. \textbf{P8} described exploring machine learning to identify zoonotic risks associated with wildlife trade. \textbf{P14} described using AI for rapid overviews and assessments of upcoming or existing crises, while \textbf{P15} described exploratory computer-vision applications for demining using satellite imagery and drone analysis.

\begin{quote}\emph{
``If we can detect signals earlier, we can respond earlier. That could change how humanitarian response works'' \textbf{-P14}
}\end{quote}

Participants emphasized that AI can amplify the results of missions by promoting new forms of collaboration between different experts and AI researchers, which can be achieved by developing different models that are specific to different domains, strengthening partnerships with academic institutions, and developing evaluation frameworks that link AI results with programmatic impact.

Overall, the participants' narratives suggest they view AI not only as an administrative tool but also as a way to amplify mission results. This can be achieved by promoting collaboration between experts and AI researchers, developing models specific to different domains, strengthening partnerships with academic institutions, and creating evaluation frameworks that link AI results with programmatic impact.

\subsubsection{Human-Centered and Collaborative Futures}\label{subsec:rq3_humancentered}

Participants saw AI as a kind of supportive layer designed to complement, rather than replace, human expertise. They underlined the importance of automating routine tasks such as documentation and the processing of information, while keeping significant decisions within human control—especially in situations that involve ethical considerations, the need for contextual interpretation, or engagement with the community \textbf{P4, P7, P8, P9, P11, P12, P13}. A number of them also highlighted the importance of incorporating AI into daily work practices through ongoing learning and by forming partnerships with academic bodies, technical organizations, and other peer NGOs. From these perspectives, we can indicate that participants envision the future use of AI as human-centered, collaborative, and organizationally embedded rather than as standalone automation.

\section{Discussion}

\subsection{When Organizational Values Shape AI Adoption Decisions}

One implication of our research is a question that previous studies into technology adoption in MDOs have not adequately addressed, \textit{what happens when there is a clash between AI adoption and the values that underlie the organizational mission?} Participants did not describe these tensions as simple trade-offs between efficiency and ethics; instead, when a use case for AI was identified as potentially undermining a fundamental principle, such as environmental sustainability, humanitarian neutrality, or community trust, they described these concerns as shaping whether and under what conditions AI should be used.

Environmental practitioners pointed out that AI applications that did not match sustainability objectives were considered computationally too complex; in addition, humanitarian workers refrained from utilizing tools of analysis that posed ethical issues in political settings. As shown in RQ2~\S\ref{subsec:rq2}, these concerns sometimes led participants to exclude particular applications or restrict their use. This pattern differs from commercial AI adoption, which is often evaluated in terms of business objectives, feasibility, and expected value, while private-sector AI ethics tends to focus on customer-facing and technically addressable concerns\cite{davenport2018, schiff2021ai}.

In MDO settings, neutrality, accountability, and the building of community trust can function as boundary conditions  \cite{goldkind2025ai, bopp2017disempowered}. Existing adoption models such as Technology Acceptance Model (TAM) \cite{davis1989usefulness} and Unified Theory of Acceptance and Use of Technology (UTAUT) \cite{venkatesh2003unified} explain individual technology acceptance and therefore provide limited insight into MDO contexts where mission values shape whether and under what conditions adoption proceeds.

Prior work on technology adoption in MDOs has shown that organizations are often pushed toward tools that tend to be directed towards technologies that promote upward accountability to the donors rather than downward accountability to the communities \cite{bopp2017disempowered, voida2011shapeshifters}. Our findings extend this argument in an important direction; even under donor pressure, participants described mission commitments as shaping whether AI was adopted, restricted, or excluded. Values, therefore, become conditions that shape whether or not an adoption is feasible and not just aspects for consideration alongside efficiencies.  As \citet{lin2024come} show, AI4SG projects typically align with the preferences of donors and developers rather than the organizations meant to benefit from such technologies, resulting in tools that rarely reach deployment. Our findings reinforce this by showing that responsible AI deployment in mission-driven contexts requires technology ecosystems built around organizational value boundaries from the start.

\textbf{Implications for HCI.} In communities that are already socially excluded due to geographic, linguistic, or socio-economic factors, AI systems that do not account for these boundaries do not just fail to help; they risk reinforcing the exclusions that MDOs exist to address. When civil society organizations dependent on vendor infrastructure cannot guarantee how beneficiary data is used, the populations most in need of protection bear the greatest risk. From an HCI perspective, these risks emphasize the necessity of designing human-AI systems in which organizations can define what tasks may be delegated to AI and where human judgment and accountability must be preserved. Further, another emerging trend highlighted by the findings is that even if organizations recognize these boundaries and plan accordingly, there is still a significant disparity between individual usage of AI technology and organizational capacity to deploy it responsibly. This suggests that responsible human--AI collaboration goes beyond individual use to rely also on organizational mechanisms that support oversight, accountability, and control over AI-supported work.

\textbf{Recommendations for MDOs.} On the basis of these results, it would be recommended that MDOs consider performing an evaluation of potential AI applications based on their suitability to mission requirements before deployment. The evaluations would  identify applications that should proceed, remain conditional on safeguards including human oversight or data control, or be excluded where unresolved risks conflict with their values and commitments. Moreover, MDOs may also consider requiring greater transparency about AI model provenance and data processing when selecting external AI providers.

\subsection{Why Individual AI Use Does Not Scale to Organizational Practice}
Our findings surface  a trend that previous studies on the adoption of technologies by public organizations have started to uncover, individual interest in AI technology does not necessarily imply the ability of an organization to use it responsibly \cite{molinari2021ai,neumann2024exploring}. We extend this finding by showing the reasons behind the persistence of such a gap across MDOs.

Interviews showed that MDOs' staff used tools like ChatGPT consistently in their personal workflows while their organizations had still failed to leave the pilot phase behind. The gap that existed between these two situations could not be explained by a lack of awareness or willingness alone. As discussed in  RQ2~\S\ref{subsec:rq2}, staff were unable to assess vendors' claims, guarantee data compliance, make sure the systems meet  regulatory requirements, or integrate the personal experiments within organization-wide governance mechanisms. The pilots produced valuable insights, yet often did not move forward due to the lack of organizational structure required to support them, such as data structures, technical expertise, and accountability channels. This phenomenon is consistent with findings from other sectors where AI fails to scale at the organizational level \cite{haefner2023implementing, selten2024organizing, uren2023technology}.

We provide a structural explanation for this phenomenon. This gap cannot be explained because of the lack of training alone; instead, it comes from the persistent under-resourcing of MDOs. Data collection efforts have been done on a case-by-case basis to fulfill the requirements of the donors, rather than building a knowledge infrastructure. Also, there are heavy administrative duties assigned to the workers, which restrict their ability to systematically learn about AI or governance. Previous studies have suggested that a higher level of AI literacy involves not just knowledge, but also institutional ability, process design, and governance \cite{long2020ai, benlian2025ai, cetindamar2022explicating, palade2021organizational}. According to our findings, a key missing component of this broader form of AI literacy is organizational capability, including the infrastructure, expertise, and governance needed to support AI beyond individual use.

\textbf{Implications for HCI.} In the context of uneven access to and understanding of AI in MDOs, organizations working among more vulnerable communities, who usually have lower resources than other organizations, will be those that lack the ability to engage effectively with AI. In such a case, there is likely to be a divide between the two categories of organizations, where well-resourced MDOs may be better positioned to benefit from AI, while the others become even more disadvantaged in their ability to serve the community. For HCI, this suggests that responsible human--AI collaboration should account for differences in organizational technical, financial, and governance capacity. Systems that assume substantial in-house expertise or infrastructure may therefore reproduce existing inequalities in who can adopt and govern AI effectively.

\textbf{Recommendations for MDOs.} In order to move beyond  individual experimentation,  MDOs may need to invest in the organizational conditions that support sustained AI use. Such arrangements would require not only improving the standardization and integration of data but also creating clear criteria for evaluating AI technologies and creating channels through which successful individual and team-level experiments can be tested for wider applicability inside the organization.  Capacity-building initiatives should thus include not only individual AI literacy but also governance and implementation considerations.\looseness=-4

\subsection{What Moving from Pilots to Practice Would Require}
Our findings raise questions that go beyond what the existing barrier literature covers. If MDOs know why pilot programs stall, what do they think would move things forward?  The participants think a strong knowledge base, ongoing AI governance, mission-driven deployments, and human-centered teams will define the future; participants described these as important conditions for responsibly scaling up AI initiatives.

Participants generally see AI as a way to easily connect organizational knowledge across different offices, languages, and time. This idea tackles one of the main adoption barriers identified in the research, which is that data fragmented across incompatible systems, often built for donor reporting rather than organizational learning, is one of the barriers that prevents pilots from scaling. Previous work on knowledge management argues that information systems are only as effective as the knowledge infrastructure supporting them \cite{alavi2001knowledge}. As shown in RQ2~\ref{subsec:rq2}, in AI deployment in MDO contexts, knowledge infrastructure therefore becomes an important part of AI infrastructure, and weaknesses in one can constrain the sustainability of the other \cite{johnk2021ready}.

Infrastructure control was recognized as an important condition for AI adoption. The participants who envisioned sovereign, on-premises, and open-source infrastructure described a need for governance rather than a mere technical preference. As shown in RQ3–Section \S\ref{subsec:rq3_sovereignty}, participants described vendor dependency as a risk to organizational accountability and autonomy. This is connected to the larger discussion on the subject of data sovereignty and how technological dependency exacerbates power imbalances \cite{erete2025towards}. For MDOs active in sensitive political fields, lack of control over data infrastructure directly threatens the trust relationships crucial to their operations \cite{suchikova2025extending}.

Lastly, participants emphasized the importance of maintaining humans involved in decision-making for significant impacts, showing they understand where the real responsibility lies in mission-driven work. With regards to the risks posed by automated results in case of vulnerable communities or controversial data, organizations may never be prepared enough to foresee all the dangers.  The approach to human-AI collaboration, in which participants believe that AI is responsible for handling the routine tasks, and humans have control over the decisions that affect people's lives, fits within the broader discussions in responsible computing about the limitations of automation in high-stakes domains \cite{bhatnagar2025bridging}. What our findings add is that this position is not resistance to AI but a considered governance stance shaped by participants' direct experience of what goes wrong when human judgment is removed from consequential decisions.

\textbf{Implications for HCI.} The conditions for responsible scaling identified by the participants, which include sovereign infrastructure, mission-aligned knowledge systems, and human oversight of consequential decisions, may be difficult to be achieved by individual organizations working alone. These conditions imply that the MDOs have to invest in both the development of data standards and open-source tools within their own sectors, as well as funding models that view the knowledge infrastructure as a key competency of an organization, not overhead  \cite{lin2024come, godefroid2024identifying, kwet2019digital}. Without this, the gap between well-resourced and under-resourced civil society organizations will widen, and the communities most dependent on MDOs will bear the consequences. For HCI, these findings suggest that the design of human--AI systems should account for the organizational infrastructures through which AI is governed, including mechanisms for data control, human oversight, and institutional accountability.

\textbf{Recommendations for MDOs. }It could also be helpful to define explicit conditions under which AI projects can move from pilots into sustained use; for instance, these conditions could include the readiness of organizations in terms of having sufficient data maturity, identification of ownership of reviews of the results, security of the sensitive data, and sufficient controls for vendors and infrastructure. Also, documenting the learnings of the pilot testing phase could allow MDOs to keep knowledge of the context-specific failures, limitations, and governance needs rather than having to carry out similar experiments across different teams.

\section{Limitations}

Our findings should be interpreted in light of several limitations. First, our sample of 15 participants, while spanning multiple regions and sectors, is skewed toward large international NGOs in the Global North, which may limit the representation of smaller or community-based organizations. Second, as a qualitative interview study, our results provide depth rather than statistical generalizability. Third, our data capture perspectives at a single point in time (March–May 2025) in a rapidly evolving technological landscape. Fourth, our analysis is interpretive and may reflect the researcher's positionality despite efforts to ensure rigor through triangulation and participant validation.

Additional methodological details and extended discussion of limitations are provided in Appendix~\ref{appendix:limitations}.

These limitations caution against overgeneralization but do not diminish our contribution in providing a systematic account of AI adoption in mission-driven organizations under real-world constraints.

\section{Positionality} \label{sec:pos} 
Our research team combines expertise in human-computer interaction (HCI), responsible and socio-technical AI, organizational studies, and computational social science, with members from both the Global North and Global South. Participants were recruited through a professional network of 27 MDOs, including INGOs and UN agencies, with which the research team collaborates on AI-related initiatives. Network partners facilitated access to practitioners involved in AI adoption and governance. These relationships may have influenced participant selection and data interpretation.

\section{Conclusion} 
Mission-driven organizations (MDOs) frequently faced difficulties in taking their AI projects beyond the pilot stage, and the participants identified the causes of these problems as being limited internal expertise, fragmented program data, a tendency within the organization to be careful, ethical considerations, and their dependence on commercial providers. As a result, individual attempts to experiment with tools such as ChatGPT did not always lead to formal adoption by the organization. Participants assessed whether specific AI applications aligned with organizational commitments to sustainability, neutrality, and community trust. Many preferred open-source or locally hosted systems to maintain greater control over sensitive data and AI-supported processes. Participants also stressed that decisions affecting vulnerable communities should remain under human oversight. These findings show that AI adoption within MDOs is dependent on capacity, governance, structures, mission alignment,  and efficiency considerations. Future research should examine whether there are variations in these variables based on the sector, organization size, and geographical contexts. \looseness=-1

\section{AI Tool Disclosure}
Large language models (LLMs) were used only to refine grammar and wording to improve the clarity and readability of the manuscript. Interview recordings were transcribed using the automated transcription tool Granola~\cite{granola2024}, and then we used LLM removal of disfluencies (e.g., ``um,'' ``hmm'') from interview transcripts, and all quotations reported in the manuscript were checked against the original recordings. No AI tools were used for data analysis, interpretation, or generating study findings; all substantive analysis was conducted by the authors.

\bibliographystyle{ACM-Reference-Format}
\bibliography{sample-base}

\section{Appendix} 

\subsection{Extended Methodology} 
\label{appendix:method}
All interviews were conducted online through video conferencing, and each interview was conducted by two interviewers. One interviewer was the primary interviewer, while the other took notes and asked follow-up questions to ensure that all necessary information was gathered. Additionally, it ensured that the interviewee's responses were clarified in real-time. After the interviews, all interview recordings were transcribed using an automated transcription tool named Granola~\cite{granola2024}.

Two researchers coded all 15 transcripts to derive initial descriptive codes, employing the QCAmap software~\cite{mayring2014qcamap}. The codes represented surface-level meaning within the data, such as \textit{``leadership skepticism,'' ``skills shortages,``} and \textit{``desire for open-source tools.''} During group meetings, the researchers discussed the codes, moving toward interpretive conceptual themes through iterative refinement of the codes. For instance, codes \textit{``leadership skepticism''} and \textit{``process misalignment''} were grouped under the interpretive conceptual theme of \textit{``institutional inertia''}, while \textit{``skills shortages''} and \textit{``struggling to keep pace''} were grouped under \textit{``implementation gaps''}. This coding process follows the progression of semantic to latent coding, as described by Braun and Clarke~\cite {braun2006thematic}.\looseness=-1

We created a preliminary codebook as part of the thematic analysis process described above. This was then reviewed by a co-author with knowledge of qualitative research methodology and AI adoption studies (who was not able to attend the above meetings). Also, we presented our preliminary results to some of the participants to validate our analysis and further refine our understanding of the organizational settings.

\label{appendix:instrument_chart}

\begin{table*}[htbp]
\centering
\caption{Thematic analysis results: Four main themes with sub-themes and representative codes}
\Description{ This table presents the results of the thematic analysis, summarizing four main themes that structure the findings: Current AI Practices, Barriers in Adoption, Future Outlook, and Recommendations. Each theme is broken down into several sub-themes, and each sub-theme lists representative qualitative codes derived from interview data. The table is organized into three columns: the first column shows the major themes, the second lists related sub-themes, and the third provides example codes illustrating patterns observed in the data. In total, the table contains four themes, seventeen sub-themes, and multiple representative codes that reflect how participants described AI use, adoption challenges, anticipated futures, and practitioner-driven recommendations. }
\label{tab:T1}
\renewcommand{\arraystretch}{1.8} 
\small
\begin{tabular}{p{2.1cm} p{4.8cm} p{7.3cm}}
\toprule
\textbf{Themes} & \textbf{Sub-themes} & \textbf{Representative Codes} \\
\midrule
\multirow{4}{2.5cm}{\textbf{Current AI practices} }  
& Outreach and External Communication & AI for Donor Intelligence \& Engagement; AI for Outreach \& External Communication; AI for Data Accessibility \& Open Information; AI Supporting Advocacy \& Public Messaging \\
& Organizational Efficiency and Planning & AI for HR \& Administrative Processes; AI for Resource Management; AI for Daily Workflow \& Knowledge Access; AI for Organizational Efficiency \& Planning \\
& Data-Driven Insight Generation & AI for Research, Text Analysis \& Reporting; AI for Data Cleaning \& Integration; AI for Pattern Recognition in Large Datasets \\
& Mission-Critical Monitoring and Response & AI for Risk Monitoring \& Anticipation; AI for Crisis Mapping \& Humanitarian Response; AI for Combating Illegal Wildlife Trade; AI for Wildlife \& Ecosystem Monitoring; AI for Sustainability \& Green Transformation \\
\midrule
\multirow{5}{2.5cm}{\textbf{Barriers in Adoption}  } 
& Implementation Gap & Skills \& Expertise Shortages; Struggling to Keep Pace with Rapid AI Development; Over-reliance on Basic Tools; Personal Anxiety \& Job Security Concerns; Limited Collaboration Between Organizations \\
& Institutional Inertia & Leadership Skepticism \& Reluctance; Process \& Organizational Alignment Barriers; Financial Constraints \& Business Case Challenges; Difficulty Measuring Impact; Country \& Context Alignment Challenges \\
& Ethical Challenges & Trade-offs Between Benefits \& Downsides; Transparency Concerns with Algorithmic Decision-Making; Inclusion \& Representation Issues; Risk of Reinforcing Structural Biases \& Exclusion; Environmental Impact Concerns \\
& Data as Asset \& Liability & Data Access \& Availability Challenges; Lack of Standardization Across Systems; Data Privacy \& Protection Concerns; Data Security Vulnerabilities; Legal \& Regulatory Compliance Challenges \\
& External Dependencies & Over-reliance on Third-party Providers; Data Sovereignty \& Protection Concerns; Geopolitical Risks \& Dependencies \\
\midrule
\multirow{4}{2cm}{\textbf{ Future Outlook} }
& Infrastructure Renaissance & AI-powered Knowledge Base \& Chatbots; Automate End-to-End Processes; AI Pilot Projects for Smart Resource Management; AI for Real-Time Translation \& Accessibility Support \\
& Institutional Sovereignty & Need for Technically Skilled Staff; Maintain Strict Control Over AI Interactions \\
& Mission Amplification & Harness AI for Biodiversity Conservation, Climate Action \& UN SDGs; Data-Driven Advocacy for Sustainability \\
& Human-Centered Innovation & Cultivate AI-Human Synergy; Preference for Open-Source AI; AI for Strategic Foresight \& Scenario Planning \\
\midrule
\multirow{4}{2cm}{\textbf{Recommendations} } 
& Capacity Building & Build AI Literacy Through Training; Leverage AI for Tedious Tasks \\
& Strategic Partnerships & Forge Cross-Sector Partnerships; Collaborate with Media Organizations \\
& Governance \& Control & Ensure Human Oversight \& Autonomy; Safeguard Data Confidentiality \\
& Implementation & Establish AI Working Groups; Adopt Clear AI Governance Policies \\
\bottomrule
\end{tabular}
\end{table*}

\subsection{Extended Limitations}
\label{appendix:limitations}
Our study should be interpreted in light of several limitations. First, while our sample of 15 participants spans humanitarian, environmental, and development organizations across multiple regions, it does not capture the full diversity of MDOs worldwide. In particular, the majority of participants came from large international NGOs in the Global North, which may bias findings toward organizations with greater digital maturity and resources. Smaller, community-based organizations, especially in the Global South, may encounter distinct challenges that are not fully represented here. \looseness=-1

Second, our qualitative interview design provides detailed accounts of organizational AI adoption but does not support statistical generalization. Quantitative or mixed-method studies could complement our findings by measuring adoption trends at scale and testing the prevalence of the barriers we identified.  

Third, the interviews capture practitioner perspectives at a single point in time (March to May 2025), in a rapidly evolving technological landscape. Since AI tools and organizational practices are changing quickly, some barriers or opportunities we identified may shift as capabilities diffuse, regulations evolve, or new forms of infrastructure emerge. Longitudinal research is needed to track how adoption trajectories unfold over time.  

Finally, our analysis was interpretive, while we incorporated external expert review and participant validation to strengthen rigor, our positionality (see ~\S\ref{sec:pos}) may have influenced coding and interpretation. Future work could build on our findings by engaging larger, more diverse samples, triangulating with organizational documents and system deployments, and comparing across different mission sectors.  

These limitations caution against overgeneralization, but they do not diminish the value of our contribution: providing one of the first systematic, cross-sectoral accounts of how MDOs are adopting and contesting AI under conditions of structural constraint and ethical tension.

\end{document}